\magnification=1200
\hsize=15.0truecm
\vsize=24truecm
\parindent=1.0truecm
\baselineskip=0.8truecm plus 0.1truecm
\parskip=0.1truecm plus 0.01truecm

\centerline{\bf MIXED STATE GEOMETRIC PHASE FROM THOMAS ROTATIONS}

\bigskip
\bigskip
\centerline{\bf P\'eter  L\'evay }
\bigskip
\centerline {\it Department of Theoretical Physics, Institute of Physics, Technical University of Budapest}
\centerline{\it H-1521 Budapest, Hungary}
\bigskip
\bigskip

\centerline{\bf Abstract}
It is shown that Uhlmann's parallel transport of purifications along a path of mixed states
represented by $2\times 2$ density matrices is just the path ordered
product of Thomas rotations.
These rotations are invariant under hyperbolic translations inside the Bloch sphere that can be regarded as the Poincar\'e ball model of hyperbolic geometry. 
A general expression for the mixed state geometric phase for an {\it arbitrary} geodesic triangle 
in terms of the Bures fidelities is derived. 
The formula gives back the solid angle result well-known from studies of the pure state geometric phase.
It is also shown that this 
mixed state anholonomy can be reinterpreted as the pure state non-Abelian anholonomy of entangled states living in a suitable restriction of the quaternionic Hopf bundle.
In this picture Uhlmann's parallel transport is just Pancharatnam transport 
of quaternionic spinors.
\bigskip
\centerline{\bf I. Introduction}
\bigskip
\bigskip
Mixed state geometric phases as introduced by Uhlmann [1] provide a natural generalization of the well-known geometric phases [2] characterizing the geometric  properties  of unitarily or nonunitarily evolving pure states.
Recently this idea of mixed state anholonomy was reconsidered within an interferometric approach [3,4], along with an alternative formulation of mixed state phases [5] . Possible experiments for confirming the appearance of such phases has been proposed and conducted (see e.g. [6]) and their relevance in the evolution of systems subjected to decoherence  through a quantum jump approach has been stressed [7].

The simplest example of an evolving
one-qubit system giving rise to a path in the space of nondegenerate $2\times 2$ density matrices was studied by many authors. 
Uhlmann himself established a formula for the geometric phase for geodesic triangles and quadrangles drawn on the surface of a sphere of {\it constant radius} inside the Bloch ball ${\cal B}$ [8].
An explicit formula for the anholonomy along an arbitrary geodesic segment in ${\cal B}$ with respect to the Bures metric
was presented by H\"ubner [9].
In a recent paper 
for evolving systems giving rise to geodesic triangles
defined by Bloch vectors  in ${\cal B}$ of {\it fixed 
magnitude}, Slater [10] compared 
Uhlmann's geometric
phase with the interferometric approach of [5].
The aim of the present paper is to point out for such systems  an interesting connection between Uhlmann's parallel transport and the phenomenon of Thomas precession. Using this  correspondence we  present a formula valid for an {\it arbitrary} geodesic triangle inside ${\cal B}$.
Our method is motivated by previous observations of Ungar [11]
that hyperbolic geometry can be useful in describing the physical and mathematical phenomena associated with one qubit density matrices. 

The organization of the paper is as follows.
In Section II. we briefly recall the background material needed
for the definition of Uhlmann's anholonomy for mixed states. 
In Section III. using the hyperbolic parametrization of one qubit density matrices we show that Uhlmann's parallel transport can be expressed as the path ordered product of
suitably defined Thomas rotations arising from the multiplication of two hyperbolic rotations (Lorentz-boosts). Here the invariance of these rotations under the so called hyperbolic translates in the Poincar\'e ball model of hyperbolic geometry is also established.
In Section IV. we present our anholonomy formula valid for an arbitrary geodesic triangle in the space of nondegenerate one qubit density matrices.
Our result for the pure state limit gives back the solid angle rule well-known from studies concerning the geometric phase [2,5,10].
Here an error of Ref. [10] is also pointed out.
Section V. is devoted to establishing a connection between our mixed state anholonomy and the pure state non-Abelian geometric phase.
Section VI. is left for the comments and conclusions.

\eject
\bigskip
\centerline{\bf II. Mixed state anholonomy}
\bigskip
\bigskip
According to Uhlmann [1] mixed state anholonomy can be defined by
lifting the curve ${\rho}(t)$ living in the space of strictly positive density operators  
to the space of its purifications such that the representative curve  $W(t)$ of these purifications is parallel translated with respect to a suitable connection.
$W$ purifies ${\rho}$ if we have $\rho =WW^{\dagger}$ and $TrWW^{\dagger}=1$.
For the special case of purifications $W$ that are elements of ${\cal H}\otimes{\cal H}^{\ast}$  with ${\cal H}$ is a finite $n$-dimensional Hilbert space and give rise to strictly positive density matrices  we have $ W\in GL(n, {\bf C})$.
It is obvious that the process of purification is ambiguous, $W$ and $WU$ with $U\in U(n)$ gives rise to the same $\rho$. 
Hence we have a (trivial) principal bundle with total space $GL(n,{\bf C})$
base ${\cal D}_n^+\equiv GL(n, {\bf C})/U(n)$ and fiber $U(n)$.
According to the connection defined by Uhlmann two purifications $W_1$ and $W_2$ giving rise to ${\rho}_1$ and ${\rho}_2$ respectively are parallel iff

$$W_1^{\dagger}W_2=W_2^{\dagger}W_1>0.\eqno(1)$$

\noindent
Using the polar decompositions $W_1={\rho}_1^{1/2}U_1$ and,
 $W_2={\rho}_2^{1/2}U_2$, that are arising as the right translates by $U(n)$
 of the global section ${\rho}^{1/2}$ one can check that $U_1$ and $U_2$ 
 related as

$$Y_{21}\equiv U_2U_1^{\dagger}={\rho}_2^{-1/2}{\rho}_1^{-1/2}({\rho}_1^{1/2}{\rho}_2{\rho}_1^{1/2})^{1/2}\eqno(2)$$

\noindent
gives rise to purifications satisfying (1). 
Dividing our path $\rho(t)\in {\cal D}_n^+$ with $0\leq t\leq 1$ into small segments   
one gets

$$W_1W_0^{\dagger}={\lim_{m\to \infty}}X_{1s_1}X_{s_1s_2}\dots 
X_{s_{m0}}{\rho}_0,\quad X_{st}=\rho_t^{-1/2}(\rho_t^{1/2}\rho_s\rho_t^{1/2})^{1/2}\rho_t^{-1/2}\eqno(3)$$

\noindent
where "lim" indicates the process of going to finer and finer subdivisions producing a continuous  path $C\in {\cal D}_n^+$.
Although Eq. (3) with the basic building blocks beeing the $X_{st}$ was usually used in the literature, for later use we favour an alternative one expressed in terms of the $Y_{st}$ defined by (2) as

$$W_1W_0^{\dagger}=\lim_{m\to\infty}\rho_1^{1/2}Y_{1s_1}Y_{s_1s_2}\dots Y_{s_{m0}}\rho_0^{1/2}.
\eqno(4)$$

\noindent
This expression (the anholonomy of the curve $C$) will be our basic one for the description of parallel transport of purifications over a path $C\in {\cal D}_n^+$.
For a closed path $C$ we have $\rho_0=\rho_1$ hence for the trace of this expression we get

$${\rm Tr}(W_1W_0^{\dagger})=\lim_{m\to\infty}{\rm Tr}(Y_{1s_1}Y_{s_1s_2}\dots Y_{s_{m0}}\rho_0).\eqno(5)$$

\noindent
The quantity ${\Phi}_g=\arg{\rm Tr}(W_1W_0^{\dagger})$ is the generalization of the geometric phase for mixed states.
and the magnitude $\nu\equiv \vert{\rm Tr}(W_1W_0^{\dagger})\vert$ is the visibility [5].

\bigskip
\centerline{\bf III. Mixed state anholonomy as Thomas rotation}
\bigskip

Now we start discussing our main concern here, namely mixed state anholonomy for a qubit system. For these systems ${\rho}$ is an element of the interior of the usual Bloch-ball ${\cal B}$. 
Moreover, the space of purifications for strictly positive $2\times2$ density
matrices $\rho\in {\cal D}_2^+\simeq {\rm Int}{\cal B}$ is $GL(2, {\bf C})$.
Hence we have $\rho=WW^{\dagger}$ with $W\in GL(2,{\bf C})$.
It is particularly instructive to regard this space
as the space of normalized entangled states for a bipartite system i.e. to have ${\cal H}\otimes {\cal H}^{\ast}\simeq {\bf C}^2\otimes{\bf C}^2$, a description giving
rise to our $\rho$ upon taking the partial trace with respect to the second subsystem.

For this we write the entangled state $\vert \Psi\rangle\in {\bf C}^2\otimes{\bf C}^2$
in the form

$$\vert \Psi\rangle=\sum_{j,k}W_{jk}\vert jk\rangle,\quad
\vert j,k\rangle\equiv {\vert j\rangle}_1\otimes{\vert k\rangle}_2 ,\quad j,k=0,1\quad W_{jk}\equiv{1\over {\sqrt{2}}}\left(\matrix{a&b\cr c&d\cr}\right).\eqno(6)$$

\noindent
The normalization condition $\langle\Psi\vert\Psi\rangle =1$ 
in this picture corresponds to the 
constraint ${\rm Tr}(WW^{\dagger})=1$.
Moreover, calculating the trace of the pure state density matrix $\vert\Psi\rangle\langle \Psi\vert$ with respect to the second subsystem yields our $\rho$,
i.e. in the ${\vert j\rangle}_1 $ base we have  $\rho={\rm Tr}_2\vert\Psi\rangle\langle\Psi\vert=WW^{\dagger}$.
Since we are considering strictly positive density matrices we have the constraint
${\rm Det}\rho\neq 0$. In terms of the matrix elements of $W$ it means $ad-bc\neq 0$
i.e. $W\in GL(2, {\bf C})$.
It is well-known that for this bipartite system the measure of entanglement
is the {\it concurrence} {\cal C} [12] which can be written as  

$$0\leq {\cal C}\equiv\vert ad-bc\vert\leq 1.\eqno(7)$$

\noindent
Separable states corresponding to reduced density matrices with a zero eigenvalue are precisely the ones with ${\cal C}=0$.

Let us parametrize our density matrix as $\rho ={1\over 2}(I+{\bf u\sigma})=WW^{\dagger}$ with $\vert {\bf u}\vert <1$.
This means we have $2u_3 =\vert a\vert^2+\vert b\vert^2-\vert c\vert^2-\vert d\vert^2$ and $u_1+iu_2=\overline{a}c+\overline{b}d$, and
it is easy to check that
${\cal C}=\sqrt{1-\vert {\bf u}\vert ^2}=4{\rm Det}\rho$.
The four real numbers $-1<u_1,u_2,u_3< 1$ and $0<u_4\equiv{\cal C}\leq 1$ can be regarded as coordinates on the upper hemisphere of a three-dimensional sphere $S^3$
embedded in ${\bf R}^4$ homeomorphic to ${\rm Int}{\cal B}$. 

In the following it is convenient to introduce a new (hyperbolic) parametrization
for $\rho$ by introducing the rapidities [13,14] ${\theta}$ as

$$\vert{\bf u}\vert =\tanh {\theta}_u,\quad 0\leq {\theta}_u < \infty.\eqno(8)$$

\noindent
In this parametrization the concurrence is related to the quantity ${\gamma}_u\equiv \cosh\theta_u$ of special relativity as
${\cal C}={\gamma}_u^{-1}$.
The reader can verify that in this case

$${\rho}^{1/2} =\sqrt{{{\cal C}\over 2}}\left(\cosh{\theta_u\over 2}I+
\sinh{\theta_u\over 2}\hat {\bf u}\sigma\right)\equiv \sqrt{{\cal C}\over 2}L(\theta_u, \hat{\bf u}),\quad \hat {\bf u}\equiv {{\bf u}\over {\vert{\bf u}\vert}},\eqno(9)$$

\noindent
where $L(\theta_u, \hat{\bf u})$ is a Lorentz boost in the spinor representation.

Using the relation $M^2-M{\rm Tr M}+{\rm Det}M=0$ (valid for $2\times 2$ matrices) and its trace we have the formula

$$\sqrt{M}={M+\sqrt{{\rm Det}M}\over {\sqrt{{\rm Tr}M+2\sqrt{{\rm Det}M}}}}.\eqno(10)$$

\noindent
Now we calculate the quantity $Y_{uv}$  of (2) with the density matrices ${\rho}_u$ and
${\rho}_v$. For this we insert $M\equiv \rho_v^{1/2}\rho_u\rho_v^{1/2}$ in Eq. (10). 
First we note that the square  of the denominator of this equation has the form

$${\rm Tr}(\rho_u\rho_v)+2\sqrt{{\rm Det}(\rho_u\rho_v)}={1\over 2}(1+{\bf uv}+{\cal C}_u{\cal C}_v)=F({\bf u},{\bf v}),\eqno(11)$$

\noindent
where $F({\bf u},{\bf v})$ is the Bures fidelity.
Moreover, by virtue of Eq. (9) its enumerator multiplied by $\rho_u^{-1/2}\rho_v^{-1/2}$ has the form

$$ \rho_u^{1/2}\rho_v^{1/2}+{{\cal C}_u{\cal C}_v\over 4}\rho_u^{-1/2}\rho_v^{-1/2}={1\over 2}\sqrt{{\cal C}_u{\cal C}_v}\left(L(\theta_u,\hat{\bf u})L(\theta_v, \hat{\bf v})+L(\theta_u,-\hat{\bf u})L(\theta_v, -\hat{\bf v})\right).\eqno(12)$$

\noindent
Recalling that the composition of two boosts can be written as another boost times a Thomas rotation the right hand side of Eq. (12) can be written as

$$L(\theta_w,\hat{\bf w})R(\alpha, \hat {\bf n})+L(\theta_w,-\hat{\bf w})
R(\alpha,{\bf n})=2\cosh{\theta_w\over 2}R(\alpha, \hat{\bf n})\eqno(13)$$

\noindent
where $R(\alpha, \hat{\bf n})=\cos{\alpha\over 2} I+i\sin{\alpha\over 2} {\bf \sigma \hat{n}}$
is the Thomas rotation matrix in the spinor representation.
By writing out explicitly the product of two boosts in the spinor representation the dependence of the quantities $\alpha$ , $\theta_w$, ${\bf w,n}$ 
on the original ones ${\theta}_{u}$, $\theta_v$  and ${\bf u, v}$ can be established  (see e.g.[15] and references therein).
One of such formulas we need is the hyperbolic law of cosines [13,15] which can be written as

$$\cosh\theta_w=\cosh\theta_u\cosh\theta_v+\sinh\theta_u\sinh\theta_v \hat{\bf u}\hat{\bf v}={1\over {\cal C}_u{\cal C}_v}\left(1+\bf {uv}\right).\eqno(14)$$

\noindent
Using this we have $2\cosh{\theta_w\over 2}=2\sqrt{F({\bf u}, {\bf v})\over {\cal C}_u{\cal C}_v}$. Putting this into Eqs. (12-13) and using Eq. (11) we obtain our result

$$Y_{uv}=\rho_u^{-1/2}\rho_v^{-1/2}(\rho_v^{1/2}\rho_u\rho_v^{1/2})^{1/2}=
R(\alpha, \hat{\bf n}).\eqno(15)$$ 

\noindent
Hence according to Eq. (4) Uhlmann's parallel transport can be understood as a sequence of Thomas rotations.

For the calculation of the explicit form of $R(\alpha, \hat{\bf n})$
we use the left hand side of Eq.(12) and the explicit forms of $\rho_u^{1/2}$
and $\rho_v^{1/2}$
obtained from Eq. (9) by expressing the hyperbolic functions in terms of
${\cal C}_u$, and ${\cal C}_v$.

$$\rho_{u}^{1/2}={1\over 2\sqrt{1+{\cal C}_u}}(1+{\cal C}_u+{\bf u\sigma}),
\quad
\rho_{v}^{1/2}={1\over 2\sqrt{1+{\cal C}_v}}\left(1+{\cal C}_v+{\bf v\sigma}
\right).\eqno(16)$$

\noindent
Collecting everything we get 

$$ Y_{uv}=R(\alpha, \hat{\bf n})=\cos{\alpha\over 2}I+i\sin{\alpha\over 2}\hat{\bf n}{\bf \sigma}=
{[(1+{\cal C}_u)(1+{\cal C}_v)+{\bf uv}]I+i({\bf u}\times{\bf v})\sigma
\over{\sqrt{[(1+{\cal C}_u)(1+{\cal C}_v)+{\bf uv}]^2+\vert\bf{u}\times{\bf v}\vert^2}}}.\eqno(17)$$

\noindent
(Compare this explicit formula with the implicit one of Eq. (2) of Ref. [9].)
From this

$$\tan{\alpha\over 2}={\vert{\bf u}\times{\bf v}\vert\over {(1+{\cal C}_u)(1+{\cal C}_v)+{\bf uv}}},\quad \hat{\bf n}={{\bf u}\times {\bf v}\over {\vert{\bf u}\times{\bf v}\vert}}.\eqno(18)$$

According to [1] this parallel transport is the one occurring along the shortest geodesic with respect to the Bures metric between the two points ${\bf u}$ and ${\bf v}$ in the interior of the Bloch-ball.
Since every smooth curve can be approximated by a sequence of geodesic segments
Eq. (4) can be regarded as such an approximation.
According to our result the parallel transport of purifications along a smooth curve in ${\cal B}$ can be represented as the path ordered pruduct of Thomas rotations.

Let us examine the (17) expression for the anholonomy transformation matrix.
First we introduce a new parametrization

$${\bf a}\equiv {{\bf u}\over{1+{\cal C}_u}}=\tanh{\theta_u\over 2}\hat{\bf u}, \quad 
{\bf b}\equiv {{\bf v}\over{1+{\cal C}_v}}=\tanh{\theta_v\over 2}\hat{\bf v}.\eqno(19)$$

\noindent
It is clear that ${\bf a}$ and ${\bf b}$ are still elements of the Bloch-ball,
they are of the same direction but different length.
In terms of these new variables $R(\alpha,\hat{\bf n})$ can be written as

$$R(\alpha, \hat{\bf n})\equiv R({\bf a}, {\bf b})={(1+{\bf ab})I+i({\bf a}\times{\bf b}){\bf \sigma}\over
\sqrt{1+2{\bf ab}+\vert {\bf a}\vert^2\vert{\bf b}\vert^2}}\in SU(2),\eqno(20)$$

\noindent
where we have used the formula $\vert{\bf a}\times{\bf b}\vert^2=\vert{\bf a}\vert^2\vert{\bf b}\vert^2-({\bf ab})^2$.
Notice that in this notation the Bures fidelity is related to the square of the denominator of this formula via the identity

$$F({\bf u},{\bf v})={1\over 4}(1+{\cal C}_u)(1+{\cal C}_v)(1+2{\bf ab}+\vert{\bf a}\vert^2\vert{\bf b}\vert^2)
=1-{\vert{\bf a}-{\bf b}\vert^2\over{(1+\vert{\bf a}\vert^2)
(1+\vert{\bf b}\vert^2)}}.\eqno(21)$$

\noindent
Here the second equality of Eq.(21) also reveals the relationship of $0\leq F({\bf u}, {\bf v})< 1$
to the distance on the upper hemisphere of $S^3$, as can be checked by stereographic projection from the south pole of $S^3$ to ${\bf R}^3$ that maps the upper hemisphere of $S^3$ to ${\cal B}$.
An alternative form of (21) is $F({\bf u}, {\bf v})={\cos}^2{\Delta \over 2}$,
where $0\leq\Delta\leq \pi$ is the geodesic distance between ${\bf a}$ and ${\bf b}$ with respect to the metric on ${\cal B}$ arising via this stereographic projection.  

Now let us define the hyperbolic-translation [16] of the vector ${\bf b}$
by the vector ${\bf a}$
as

$${\tau}_{\bf a}({\bf b})\equiv {(1-\vert{\bf a}\vert^2){\bf b}+(1+2{\bf ab}+\vert{\bf b}\vert^2){\bf a}\over {1+2{\bf ab}+\vert{\bf a}\vert^2\vert{\bf b}\vert^2}}.\eqno(22)$$

\noindent
Denoting ${\bf a^{\prime}}\equiv{\tau}_{\bf a}({\bf b})$ one can show that $\vert{\bf a^{\prime}}\vert^2=\vert{\bf a}+{\bf b}\vert^2/(1+2{\bf ab}+\vert{\bf a}\vert^2\vert{\bf b}\vert^2)\leq 1$ i.e. this transformation maps ${\cal B}$ onto itself.
Using this we get $(1+2{\bf a^{\prime}b}+\vert{\bf a^{\prime}}\vert^2\vert{\bf b}\vert^2)(1+2{\bf ab}+\vert{\bf a}\vert^2\vert{\bf b}\vert^2)=(1+2{\bf ab}+\vert{\bf b}\vert^2)^2$.
Now it is easy to establish the formula

$$R({\bf a},{\bf b})=R(\tau_{\bf a}({\bf b}),{\bf b})=
R({\bf a},\tau_{\bf b}({\bf a})).\eqno(23)$$

\noindent
Equation (23) states the important result that the anholonomy  
for Uhlmann's parallel transport is invariant with respect to hyperbolic-translations of the Bloch-ball regarded as the Poincar\'e ball modell of hyperbolic geometry. These properties were called by Ungar [11]  "left-loop" and "right-loop" properties in his study of  density matrices and gyrovector spaces.
In this way we established an implementation of his abstract setting up on Uhlmann's parallel transport of purifications.
We note in closing that the (23) properties can be used to find deformations of
curves consisting of geodesic segments having the same anholonomy, a property
that can be useful for the experimental verification of Uhlmann parallelism in this most general setting up.

\bigskip
\centerline{\bf IV. Geodesic triangles}
\bigskip

In this section we use the results of the previous section to
derive an explicit formula for the anholonomy of a special closed path: the geodesic triangle.
Note, that this problem has already been considered in Refs. [8] and [10]
for the three points of the triangle lying on a spherical shell of ${\cal B}$
of {\it fixed concurrence}.
Here we consider the general case and chose three {\it arbitrary} points ${\bf u}$, ${\bf v}$ and ${\bf w}$ in the interior of the Bloch-ball. The corresponding concurrences are ${\cal C}_u$,
${\cal C}_v$ and ${\cal C}_w$. We renormalize our vectors ${\bf u}$, ${\bf v}$
and ${\bf w}$ as in Eq. (19), the resulting vectors still belonging to ${\rm Int}{\cal B}$ are ${\bf a}$, ${\bf b}$ and ${\bf c}$.
According to Eqs. (4,17,20) in order to calculate the anholonomy we have to evaluate the quantity

$${\cal R}({\bf a}, {\bf b}, {\bf c})\equiv R({\bf a}, {\bf c})R({\bf c}, {\bf b})R({\bf b}, {\bf a})=\cos{\delta\over 2}
+i\sin{\delta\over 2}{\bf m}{\bf \sigma},\eqno(24)
$$

\noindent
where the last equality expresses the fact that the resulting matrix should
also have an $SU(2)$ form. 

By virtue of Eqs. (20-21) we can extract a factor from ${\cal R}({\bf a}, {\bf b}, {\bf c})$  of the form
${1\over 8}(1+{\cal C}_u)(1+{\cal C}_v)(1+{\cal C}_w)/\sqrt{F({\bf u}, {\bf v})
F({\bf w},{\bf v})F({\bf v}, {\bf u})}$.
Hence we merely have to evaluate the matrix

$$J\equiv (I+a\cdot c)(I+c\cdot b)(I+ b\cdot a),\quad {\rm where} \quad
a\cdot c\equiv ({\bf a\sigma})({\bf c\sigma})= ({\bf ac})I+i({\bf a}\times{\bf c})\sigma\quad {\rm etc.}\quad\eqno(25)$$

\noindent
Straightforward calculation yields the result

$$J=\left(1+a^2b^2c^2+(1+a^2)({\bf bc})+(1+b^2)({\bf ca})+(1+c^2)({\bf ab})\right)I$$

$$
+[c-({\bf ca})a,b-({\bf ab})a]
+{1\over 2}\left((1-a^2)[b,c]-(1-b^2)[c,a]-(1-c^2)[a,b]\right),\eqno(26)$$

\noindent
where $a\equiv ({\bf a\sigma})$, $a^2\equiv\vert{\bf a}\vert^2$ etc. and $[a,c]$ denotes the commutator of the corresponding matrices.
By virtue of the relations $a^2=(1-{\cal C}_u)/(1+{\cal C}_u)$ etc. and the (11) definition of the Bures fidelity we obtain for that part of ${\cal R}({\bf a}, {\bf b}, {\bf c})$
 which is proportional to the identity matrix
the formula

$$\cos{\delta\over 2}={F({\bf u},{\bf w})+F({\bf w}, {\bf v})+F({\bf v},{\bf u})-1\over 2\sqrt{F({\bf u}, {\bf w})F({\bf w}, {\bf v})F({\bf v}, {\bf u})}}.\eqno(27)$$

In order to also find the axis of rotation we introduce the vectors

$${\bf p}\equiv { {(1+a^2){\bf c}-(1+2{\bf ac}-c^2){\bf a}}\over {1+2{\bf ac}
+a^2c^2}},\quad
{\bf q}\equiv { {(1+a^2){\bf b}-(1+2{\bf ab}-b^2){\bf a}}\over {1+2{\bf ab}
+a^2b^2}}.\eqno(28)$$

\noindent
With these vectors it is straightforward to check that

$${\cal R}({\bf a},{\bf b}, {\bf c})={ {(1+{\bf pq})I+i({\bf p}\times{\bf q})\sigma}\over \sqrt{1+2{\bf pq}+p^2q^2}},\eqno(29)$$

\noindent
hence the angle $\delta$ and axis ${\bf m}$ of the resulting Thomas rotation is
given by the expressions

$$\tan {\delta \over 2}={{\vert{\bf p}\times{\bf q}\vert}\over {(1+{\bf pq})}},
\quad {\bf m}={{\bf p}\times{\bf q}\over {\vert{\bf p}\times{\bf q}\vert}}.
\eqno(30)$$

A comparison of Eq. (28) and (22)
shows that the formulae for ${\bf p}$ and ${\bf q}$  up to some crucial sign changes look like the hyperbolic translates.
${\bf p}$ and ${\bf q}$ are "translates" by ${\bf a}$ of ${\bf c}$ and ${\bf b}$.  
However, by virtue of the relations
$p^2=\vert{\bf a}-{\bf c}\vert^2/(1+2{\bf ac}+a^2c^2)$, and
$q^2=\vert{\bf a}-{\bf b}\vert^2/(1+2{\bf ab}+a^2b^2)$ one can see that these "spherical translates" are not mapping ${\cal B}$ (homeomorphic to the upper half of $S^3$) onto itself. These are rather isometries of the {\it full} $S^3$ with its metric given by Eq. (21).

In order to gain some  more insight into the geometric meaning of the (28)
"spherical translate" we notice that

$${{\bf p}\over {\vert {\bf p}\vert ^2}}={\bf a}+{(1+a^2)\over {\vert {\bf c}-{\bf a}\vert ^2}}({\bf c}-{\bf a}).\eqno(31)$$

\noindent
Now recall [16] the definition of the transformation ${\sigma}^r_{\bf a}: {\bf R}^3\to {\bf R}^3$

$${\sigma}^r_{\bf a}({\bf x})\equiv {\bf a}+{r^2\over {\vert {\bf x}-{\bf a}\vert ^2}}({\bf x}-{\bf a}),\eqno(32)$$

\noindent
which is the inversion with respect to a sphere $S^2$ in ${\bf R}^3$
centered at ${\bf a}$ with radius $r$.
It is now obvious that (31) is an inversion of ${\bf c}$ with respect to a sphere centered at ${\bf a}$ with radius $r^2=1+a^2$.
Moreover, since the transformation ${\bf p}\to {\bf p}/p^2$ is also an inversion with respect to the sphere centered at the origin with radius $1$, we obtain the following result.  ${\bf p}$ (resp. ${\bf q}$) is the result of two inversions  
applied to the point ${\bf c}$ (resp. ${\bf b}$). One of the inversions is defined by the point ${\bf a}$, the point we have chosen as the starting one for the traversal of the geodesic triangle.

Now let us calculate the geometric phase corresponding to our geodesic triangle! 
First we notice that

$$1+2{\bf pq}+p^2q^2=(1+a^2)^2{1+2{\bf bc}+b^2c^2\over {(1+2{\bf ab}+a^2b^2)
(1+2{\bf ac}+a^2c^2)}}.\eqno(33)$$

\noindent
From Eqs. (5) and (24) we get  
${\rm Tr}({\cal R}({\bf u}, {\bf v}, {\bf w}){\rho}_u)=\cos{\delta\over 2}+
i\sin{\delta\over 2}({\bf um})$
hence after recalling Eqs. (19), (21), (27-30) and (33) straightforward calculation yields the result

$$\nu e^{i{\Phi}_g}\equiv {\rm Tr}({\cal R}{\rho}_u)={ F({\bf u}, {\bf v})+F({\bf v},{\bf w})+F({\bf w}, {\bf u})-1 -{i\over 2}{\bf u}({\bf v}\times{\bf w})\over
2\sqrt{F({\bf u}, {\bf v})F({\bf v}, {\bf w})F({\bf w}, {\bf u})}}.\eqno(34)$$

\noindent
Hence our formula for the mixed state geometric phase
takes the form

$$tan{\Phi}_g=-{{1\over 2}{\bf u}({\bf v}\times {\bf w})\over
{F({\bf u}, {\bf v})+F({\bf v}, {\bf w})+F({\bf w}, {\bf u})-1}}.\eqno(35)$$

\noindent
In the pure state limit we have ${\cal C}_u={\cal C}_v={\cal C}_w=0$.
Since the vectors ${\bf u}$, ${\bf v}$ and ${\bf w}$ are of unit norm
we denote them in this special case as ${\bf n}_1$, ${\bf n}_2$ and ${\bf n}_3$.
Recalling  Eq. (11) for the Bures-fidelity from (35) we obtain the result

$$\tan{\Phi}_g=-\tan {\Omega\over 2}=- {{\bf n}_1({\bf n}_2\times {\bf n}_3)\over {1+{\bf n}_1{\bf n}_2+{\bf n}_2{\bf n}_3+{\bf n}_3{\bf n}_1}}\eqno(36)$$

\noindent
which is the formula for the tangent of  minus half the solid angle [17]  corresponding to the geodesic triangle on the surface of the unit sphere $S^2$ i.e.
we have ${\Phi}_g=-{\Omega\over 2}$.
Note that a more familiar form for ${{\Omega}\over 2}$ is given by [17]

$$\cos{\Omega\over 2}={ \cos ^2 {{\theta}_{12}\over 2}+
\cos ^2 { {\theta}_{23}\over 2}+
\cos ^2 { {\theta}_{31}\over 2}-1\over {
2{\cos{\theta}_{12}\over 2}{\cos{\theta}_{23}\over 2}{\cos{\theta}_{31}\over 2}}}={1+{\bf n}_1{\bf n}_2+{\bf n}_2{\bf n}_3+{\bf n}_3{\bf n}_1\over \sqrt{2(1+{\bf n}_1{\bf n}_2)(1+{\bf n}_2{\bf n}_3)(1+{\bf n}_3{\bf n}_1)}},\eqno(37)$$

\noindent
with ${\bf n}_i{\bf n}_j\equiv \cos{\theta}_{ij}$. 
Comparing this with the pure state limit of Eq. (27) we see that in this case
the Thomas rotation angle is just the solid angle, i.e. $\delta=\Omega$.
Hence for pure states we get back to the results well-known from studies concerning the ordinary geometric phase.

Eq. (34) is the most general formula that defines the visibility $\nu$
and the Uhlmann mixed state geometric phase ${\Phi}_g$ valid for an {\it arbitrary} geodesic triangle defined by the points ${\bf u,v,w}$ inside the Bloch ball ${\cal B}$. The usual geometric phase is obtained in the limiting case of sending all of the points to the boundary of ${\cal B}$ representing pure states.
As a further investigation of formula (34) let us now consider the important special case studied by Slater [10]
when $\vert{\bf u}\vert=\vert{\bf v}\vert=\vert{\bf w}\vert =r={\rm const}$!
Let ${\bf u}\equiv r{\bf n}_1$, ${\bf v}\equiv r{\bf n}_2$ and ${\bf w}=r{\bf n}_3$, then $F({\bf u}, {\bf v})=1+{1\over 2}r^2({\bf n}_1{\bf n}_2-1)$ etc.
By virtue of (35) we have the result

$$\tan {\Phi}_g =-{r^3{\bf n}_1({\bf n}_2\times{\bf n}_3)\over
{4(1-r^2)+r^2(1+{\bf n}_1{\bf n}_2+{\bf n}_2{\bf n}_3+{\bf n}_3{\bf n}_1)}}=
-{r^3\mu\over
{4(1-r^2)+r^2\mu}}
\tan{\Omega \over 2},\eqno(38)$$

\noindent
where $\mu\equiv 1+{\bf n}_1{\bf n}_2+{\bf n}_2{\bf n}_3+{\bf n}_3{\bf n}_1$
a notation used in Ref [10]. 
This result is in contrast vith the claim of Slater (see Eq. (18) of Ref. [10])

$$\tan{\Phi}_g^{\rm Slater}=-{r^3\mu\over {4+(\mu -10)r^2+6r^4}}\tan {\Omega\over 2}.\eqno(39)$$

\noindent
Notice that after the replacement $6r^4\to 6r^2$ his result would reproduce 
the correct one of Eq.(38).  

Using ideas of interferometry an alternative definition (different from the one as given by Uhlmann) for the mixed state geometric phase appeared in Ref. [5].
In this approach the result for the situation studied above is [5]

$$\tan{\Phi}_g^{\rm int}=-r\tan{\Omega\over 2}.\eqno(40)$$

\noindent
From Eqs. (38) and (40) we see that the ratio $\tan{\Phi}_g/\tan{\Phi}_g^{\rm int}$is $r^2\mu/r^2\mu+4(1-r^2)$ hence the the two different types of mixed state phases are equal merely in the pure state ($r=1$) case.
(In Ref. [10] it was claimed that the two phases are equal also for the nontrivial case with $r=\sqrt{2/3}$, a possibility clearly following from the erroneous result of  Eq. (39).)
The fact that the two approaches give different results for the mixed state anholonomy was first stressed
in Refs. [10] and [4].

Closing this section we check that the formula for the visibility
(i.e. the magnitude of the rhs. of Eq. (34) ) gives the result $\nu=1$
in the pure state limit.
The visibility is

$$\nu=\sqrt{ (F({\bf u}, {\bf v})+F({\bf v}, {\bf w})+F({\bf w}, {\bf u})-1)^2+{1\over 4} V^2\over {4F({\bf u}, {\bf v})F({\bf v}, {\bf w})F({\bf w}, {\bf u})}},\eqno(41)$$

\noindent
where $V={\bf u}({\bf v}\times {\bf w})$ is the volume of the parallelepiped
spanned by the triple ${\bf u}$, ${\bf v}$ and ${\bf w}$ .
In the pure state limit we get

$$\nu=\sqrt{{\mu}^2+{V}^2\over {2(1+{\bf n}_1{\bf n}_2)(1+{\bf n}_2{\bf n}_3)(1+{\bf n}_3{\bf n}_1)}}.\eqno(42)$$

\noindent
Using the First Law of Cosines [16] 
$\cos {\theta}_{12}-\cos{\theta}_{23}\cos{\theta}_{31}=\sin{\theta}_{23}\sin{\theta}_{31}\cos\gamma$ of spherical trigonometry, where $\gamma$ is the angle of the spherical triangle at
the point defined by  ${\bf n}_3$,  and the relations
${\bf n}_i{\bf n}_j=\cos{\theta}_{ij}$ 
and $\sin{\theta}_{23}\sin{\theta}_{31}\sin\gamma=V$ straightforward calculation yields the expected result $\nu =1$.
Hence, for the mixed state case $\delta$ is different from ${\Omega}$ and $\nu\neq 1$
(${\bf um}\neq 1$) properties also shown by analysing the alternative formula
$\nu=\sqrt{\cos^2{\delta/2}+\sin^2{\delta/2}({\bf um})^2}$ to be compared for $r={\rm constant}$ with
formula (26) of Ref [5].

\bigskip
\centerline{\bf V. 
Mixed state anholonomy and quaternionic phases}
\bigskip

Representing the space of purifications as the space of normalized entangled states  in ${\bf C}^2\otimes{\bf C}^2$ (see Eq. (6))
we have the possibility to build up a dictionary between the nomenclatures of the  mixed and the pure state anholonomies.
In order to do this recall that due to the constraint $\langle\Psi\vert\Psi\rangle ={\rm Tr} WW^{\dagger}=1$ the space of such purifications is the seven sphere $S^7$.
Let us parametrize the matrix $W$ of (6) as

$$W={1\over {\sqrt{2}}}(Q_0+iQ_1),\quad Q_0={\alpha}_0I-i{\alpha}_j{\sigma}_j
,\quad Q_1\equiv {\beta}_0 I-i{\beta}_j{\sigma}_j.\eqno (43)$$

\noindent
(Summation over repeated indices is understood, ${\sigma}_j$ $j=1,2,3$ 
are the Pauli matrices.)
Notice that Eq. (43) amounts to a change of parametrization 
from the {\it four complex} numbers $a,b,c,d$  of Eq. (6) to the {\it eight real } ones
${\alpha}_{\mu}$ and ${\beta}_{\mu}$ $\mu =0,1,2,3$.
Explicitly we have

$$a=({\alpha}_0+{\beta}_3) +i({\beta}_0-{\alpha}_3),\quad
  b=({\beta}_1-{\alpha}_2) -i({\alpha}_1+{\beta}_2)$$
 
$$ c=({\beta}_1+{\alpha}_2)-i({\alpha}_1-{\beta}_2),\quad
   d=({\alpha}_0-{\beta}_3)+i({\beta}_0+{\alpha}_3).\eqno(44)$$

\noindent
and

$$2{\alpha}_0={\Re}a+{\Re}d,\quad 2{\alpha}_1=-{\Im}b-{\Im}c,\quad 2{\alpha}_2={\Re}c-{\Re }b,\quad 2{\alpha}_3={\Im}d-{\Im a}$$

$$2{\beta}_0={\Im}a+{\Im}d,\quad 2{\beta}_1={\Re}b+{\Re}c,\quad 2{\beta}_2={\Im}c-{\Im }b,\quad 2{\beta}_3={\Re }a-{\Re}d,\eqno(45)$$

\noindent
where the symbols ${\Re}$ and ${\Im}$ refer to the real and imaginary parts of the corresponding complex numbers.
Notice moreover, that the correspondence

$${\bf i}\leftrightarrow -i{\sigma}_1,\quad{\bf j}\leftrightarrow -i{\sigma}_2,\quad {\bf k}\leftrightarrow -i{\sigma}_3\eqno(46)$$

\noindent
defines a mapping between a $W\in GL(2, {\bf C})$ and a {\it quaternionic spinor}
$(q_0, q_1)^T \in {\bf H}^2$ i.e. we have the correspondence

$$  W\mapsto \left(\matrix{q_0\cr q_1\cr}\right),\quad q_0={\alpha}_0+{\alpha}_1{\bf i}+{\alpha}_2{\bf j}+{\alpha}_3{\bf k}, \quad q_1={\beta}_0+{\beta}_1{\bf i}+{\beta}_2{\bf j}+{\beta}_3{\bf k}.\eqno(47)$$

\noindent
On the space of two component quaternionic spinors we can define an inner product $\langle\vert\rangle : {\bf H}^2\times {\bf H}^2\to {\bf H}$
as $\langle q\vert p\rangle \equiv \overline{q_0}p_0+\overline{q_1}p_1$, i.e.
we have quaternionic conjugation in the first factor.
From the normalization condition ${\rm Tr}WW^{\dagger}=1$ it follows that
${\alpha}_\mu{\alpha}_{\mu}+{\beta}_{\mu}{\beta}_{\mu}=1$, i.e. the spinor
$(q_0, q_1)^T$ is normalized, $ {\vert\vert q\vert\vert}^2\equiv \langle q\vert q\rangle =1$.
It means that ${\alpha}_{\mu}$ and ${\beta}_{\mu}$ are Cartesian coordinates for the seven-sphere $S^7$. 

Let us express our reduced density matrix ${\rho}={\rm Tr}_2\vert \Psi\rangle\langle\Psi\vert=WW^{\dagger}$ in terms of the matrices $Q_0$ and $Q_1$
corresponding to the quaternions $q_0$ and $q_1$!
By virtue of (43) we have

$${\rho}={1\over 2}\left(Q_0Q_0^{\dagger}+Q_1Q_1^{\dagger}+i(Q_1Q_0^{\dagger}-Q_0Q_1^{\dagger})\right)={1\over 2}(I+{\bf u}{\bf \sigma})\eqno(48)$$

\noindent
where we have used the normalization condition and the fact that the matrix $Q_1Q_0^{\dagger}-Q_0Q_1^{\dagger}$ is an anti-Hermitian
$2\times 2$ one hence it can be expanded as $-i{\sigma}_ju_j$, $j=1,2,3$ with
$u_1,u_2,u_3$ are {\it real} parameters of the one-qubit density matrix. For later use we also define the quantity
$u_0\equiv 
Q_1Q_0^{\dagger}+Q_0Q_1^{\dagger}$
which is two times the Hermitian part of 
$Q_1Q_0^{\dagger}$.
The Hermitian and anti-Hermitian parts of the matrix $Q_1Q_0^{\dagger}$
correspond to the {\it real} and {\it imaginary } 
parts of the corresponding quaternion $q_1{\overline q_0}$ hence we can define the quaternion

$$u\equiv u_0+u_1{\bf i}+u_2{\bf j}+u_3{\bf k}=2q_1\overline{q_0},\quad
{\rm Re}(u)={1\over 2}(u+\overline{u})=u_0,\quad {\rm Im}(u)={1\over 2}(u-\overline{u}).\eqno(49)$$

\noindent
Let us define one more quantity

$$u_4\equiv \vert q_0\vert ^2-\vert q_1\vert ^2=Q_0Q_0^{\dagger}-Q_1Q_1^{\dagger}.\eqno(50)$$

\noindent
Recall also from Section III. that that the coordinates $u_{\hat{\mu}}$ $\hat{\mu}=0,1,2,3,4$ are related to the complex numbers $a,b,c,d$ as
$u_4+iu_0=ad-bc$, $u_1+iu_2=\overline{a}c+\overline{b}d$ and $2u_3=\vert a\vert^2+\vert b\vert^2-\vert c\vert^2-\vert d\vert^2$.
Hence the concurrence is just ${\cal C}=\sqrt{u_4^2+u_0^2}$.
It is straightforward to check that $\overline{u}u+u_4^2=u_{\hat{\mu}}u_{\hat{\mu}}=1$, hence $u_{\hat{\mu}}\in S^4$ i.e. it is an element of the {\it four dimensional} sphere.
As a result of Eqs. (49-50) one can define a map $\pi: S^7\to S^4$. 
Notice that according to the explicit form of this map the transformation
({\it right} multiplication of the quaternionic spinor with a {\it unit} quaternion)

$$\left(\matrix{q_0\cr q_1\cr}\right)\to\left(\matrix{q_0\cr q_1\cr}\right)s,\quad {\rm where}\quad \overline{s}s=1\eqno(51)$$

\noindent
leaves the coordinates $u_{\hat{\mu}}$ invariant.
Since unit quaternions correspond to elements of $SU(2)\sim S^3$
the projection $\pi$ defines a fibration (the second Hopf fibration [18]) of $S^7$ with base $S^4$ and fiber $S^3$.
Reinterpreting our quaternionic spinors as entangled states it is straightforward to show that this $SU(2)$ fiber degree of freedom corresponds to the possibility of making local unitary transformations $I\otimes S, S\in SU(2)$ in the {\it second} subsystem.
This idea of representing entanglement via the twisting of a nontrivial fiber bundle was initiated in [19] and further developed in Refs. [20] and [21]. 
Here we merely need one result from Ref. [20]: an element  $\vert q\rangle \in S^7$
can be parametrized by points of $S^4$ {\it minus the south pole} (SP) as

$$\vert q\rangle =\left(\matrix{q_0\cr q_1\cr}\right)={1\over {\sqrt{2(1+u_4)}}}
\left(\matrix{1+u_4\cr u\cr}\right)s\equiv \vert u\rangle s, \quad \overline{s}s=1,\quad u_{\hat{\mu}}\in S^4-\{ SP\}.\eqno(52)$$

\noindent
Eq. (52) is a local section of our bundle.   
There are no global sections (i.e. expressions like (52) nonsingular over all of $S^4$) which is just another way of saying that the Hopf bundle is nontrivial
i.e.  $S^7\neq S^4\times S^3$.
Of course we can define alternative sections that are singular at different points, the (52) choice is dictated by convenience.

Consider now three quaternionic spinors $\vert q\rangle =\vert u\rangle s$,
$\vert p\rangle =\vert v\rangle x$ and $\vert r\rangle=\vert w\rangle y$ $s,x,y\in SU(2)$ representing entangled states $\vert \Psi\rangle$ , $\vert \Phi\rangle$ and $\vert\chi\rangle $! Notice that the notation indicates that the corresponding quaternionic spinors are parametrized by the vectors $u_{\hat{\mu}}$ , $v_{\hat{\mu}}$ and $w_{\hat{\mu}}$ which are elements of the open neighbourhood $S^4-\{SP\}$.

As a next step we consider the trivial subbundle ${\cal E}$ of the Hopf bundle defined by the conditions $u_0=0$, and $u_4={\cal C}>0$. ${\cal E}$ is a fiber bundle with an $S^3$ fiber over the submanifold ${\cal M}$ of the upper half hemisphere of $S^4$ defined by the aforementioned constraints.
It is easy to see that ${\cal M}$ is topologically the upper half hemisphere of a three sphere defined by the coordinates ${\cal C},u_1, u_2, u_3$ and can be identified with the interior of the Bloch-ball of reduced density matrices
${\rm Int}{\cal B}$.
For more details on the structure of the bundle ${\cal E}$ that has already been studied in the context of Uhlmann's connection see Ref. [22].
Let us suppose that our spinors $\vert q\rangle$, $\vert p\rangle$ and $\vert r\rangle$ define {\it global} sections of ${\cal E}$ of the (52) form.
This means that we set the parameter values  $u_0=v_0=w_0$ in expressions like Eq. (52) to zero, and the ones $u_4,v_4,w_4$ to ${\cal C}_u ,{\cal C}_v $ and ${\cal C}_w$.

It is now straightforward to check that the unit quaternion
$\langle v\vert u\rangle/\vert\langle v\vert u\rangle\vert$ is just $Y_{vu}$
of Eq. (17). Moreover employing the notation of Eq. (19) equation (24)
can be written in the following form

$${\cal R}({\bf u}, {\bf v}, {\bf w})=s{\langle q \vert r\rangle \over {\vert\langle q\vert r\rangle\vert}}{\langle r\vert p\rangle\over {\vert\langle r\vert p\rangle\vert}}{\langle p\vert q\rangle\over {\vert \langle p\vert q\rangle\vert}}\overline{s}.\eqno(53)$$

\noindent
where it is now understood that the left hand side is also regarded as a unit quaternion.

Notice now that Eq. (52) is just the quaternionic analogue of the polar decomposition. Indeed according to Eq. (16), the spinor $\vert u\rangle$  corresponds to the matrix ${\rho}_u^{1/2}$, and the unit quaternion $s$ to the $SU(2)$ part of the $U(2)$ matrix $S$ of the polar decomposition $W_0={\rho}_u^{1/2}S$.
Since $U(2)\sim U(1)\times SU(2)$ we only have to account for a complex phase,
but  this is fixed by our choice $u_0=0$ when restricting to the subbundle ${\cal E}$. (Notice that according to Eqs. (6-7) $u_4+iu_0\equiv{\cal C}e^{i\kappa}=2{\rm Det}W$ where $\tan\kappa =u_0/u_4$. Hence the $u_0\neq 0$ case amounts to multiplying our (6) entangled state by a $U(1)$ phase.)

Now let us write Eq. (4) for the geodesic triangle in the following form

$$W_1\equiv {\Lambda}W_0={\rho}_u^{1/2}{\cal R}({\bf u}, {\bf v}, {\bf w}){\rho}_u^{-1/2}W_0,\quad {\rm where}\quad {\rho}_u=W_0W_0^{\dagger}.\eqno(54)$$

\noindent
Since the polar decomposition $W_0={\rho}_u^{1/2}S$ corresponds to the (52) section of the bundle ${\cal E}$, we can write

$$W_1= {\Lambda}{\rho}_u^{1/2}S={\rho}_u^{1/2}SS^{\dagger}{\cal R}({\bf u}, {\bf v}, {\bf w})S.\eqno(55)$$

\noindent
Using the notation $\vert q^{\prime}\rangle$ for the quaternionic representative
of $W_1$ we see that the quaternionic version of Eq. (55) is
$\vert q^{\prime}\rangle=\vert q\rangle \overline{s}{\cal R}s$.
By virtue of Eq. (53) Uhlmann's parallel transport in ${\cal E}$ in the quaternionic representation can be written as
$$\vert q^{\prime}\rangle =\vert q\rangle
{\langle q\vert r\rangle \over {\vert\langle q\vert r\rangle\vert}}
{\langle r\vert p\rangle \over {\vert\langle r\vert p\rangle\vert}}
{\langle p\vert q\rangle \over {\vert\langle p\vert q\rangle\vert}}.\eqno(56)$$

\noindent
It is clear that for an arbitrary geodesic polygon Eq. (56) has to be multiplied from the right by extra quaternionic phase factors corresponding to transitions to the new points of the polygon.
The geodesic rule obtained in this way is the non-Abelian analogue of the well-known one
obtained for filtering measurements in the context of the usual geometric phase [23], [24].
In this picture each polygon ${\Gamma}$ is decomposed into a sequence of geodesic triangles. Each triangle  ${\Delta}_j$ gives rise to a Thomas rotation of the (24) form with angle ${\delta}_j$ and axis ${\bf m}_j$.
Since ${\Gamma}$ in general is not a planar curve the rotations corresponding to different triangles have different axes. As a result the total rotation angle is not the sum of the component rotations as was in the Abelian case corresponding to the ordinary geometric phase.
In this more general case we have to combine rotations with different axis resulting in the appearance of a path ordered product.
Going to finer and finer subdivisions any smooth closed curve $C$ can be approximated by a suitable polygon ${\Gamma}$.
The resulting quaternionic phase can be written as the path ordered exponent
${\cal P}e^{-\oint_CA}$ where the $su(2)$-valued gauge-field can be written
as

$$A={\rm Im}\langle u\vert du\rangle= {1\over 2}{\rm Im}{\overline{u}du\over {1+u_4}}\quad u=u_1{\bf i}+u_2{\bf j}+u_3{\bf k},\quad u_4\equiv {\cal C}_u.\eqno(57)$$

\noindent
As was remarked in Refs. [20,22] $A$ is just the pull-back of the restriction of the canonical (instanton) connection on the quaternionic Hopf bundle to the bundle ${\cal E}$ with respect to the section $s=1$ (see Eq. (52)). 

\bigskip
\centerline{\bf  VI. Conclusions}
\bigskip

In this paper we investigated  Uhlmann's parallel transport as applied to a qubit system.
In spite of beeing the simplest and hence best studied example this system  
still shows nice geometric properties have not fully been appreciated 
by the physics community.
Our paper was intended to fill in this gap by explicitly working out these missing interesting details.
First we have shown that the very special features of the qubit system 
enable one to reinterpret Uhlmann's parallel transport as a sequence of Thomas rotations.
We have also shown some interesting connections with hyperbolic geometry.
In particular we proved that the finite Thomas rotations are invariant with respect to hyperbolic translates of the interior of the Bloch ball regarded as the Poincar\'e model of hyperbolic geometry (see Eq. (23)).
These observations should not come as a surprise since Uhlmann's parallel transport has its origin in the underlying Bures geometry [1] of the Bloch ball ${\cal B}$,
that has already been related to the Poncar\'e metric in hyperbolic geometry [13,20], moreover
it is easy to see [20] that the Bures metric is conformally equivalent to the standard Poincar\'e one. 

In section IV. we derived an explicit formula for the $SU(2)$ anholonomy 
matrix in the case of a geodesic triangle (Eqs. (28-30)).
From this an expression in terms of the Bures fidelities for the mixed state geometric phase and the visibility was derived (Eq. (35) and (41)). These general results were shown to give back in the pure state limit the corresponding ones known from studies concerning the usual geometric phase.
As far as the author knows these formulae in their full generality have not appeared in the literature yet.
The geometric significance of these expressions were elaborated, and an error that appeared in Ref. [10] was corrected.

In Section V. we managed to reformulate our results concerning the mixed state anholonomy in terms of the pure state {\it non-Abelian} one.
The idea was to reinterpret the space of purifications as the Hilbert space for an entangled two-qubit system. 
This trick enabled us to recast Uhlmann's parallel transport in yet another form i.e. in the one
of a sequence of quaternionic filtering measurements (Eq. (56)). 
By going to finer and finer subdivisions we have recovered Uhlmann's parallelism
as the Wilson loop over a gauge field which is a suitable restriction of the well-known instanton connection. 

The advantage of this quaternionic formalism is clear: Uhlmann's parallel transport
for one qubit density matrices in this representation is just the quaternionic analogue of the usual Pancharatnam transport extensively used in studies concerning the geometric phase [2, 25]. 
In this language two entangled states $\vert \Psi\rangle$ and $\vert{\Phi}\rangle$ regarded as purifications for one-qubit density matrices are "in phase" iff their quaternionic representatives $\vert q\rangle$ and $\vert p \rangle$
satisfy the constraint: $\langle p\vert q\rangle$ is real and positive.
It is easy to check that this constraint is equivalent to the one as given by Eq. (1).
Moreover, this rule provides a nice way of defining the difference of these entangled states in their local unitary transformations corresponding to the second subsystem.
Indeed, consider $\vert\Psi\rangle$ and $\vert \Phi\rangle$ as above and define their relative $U(1)$ phase to be the usual Pancharatnam phase factor  ${\langle\Phi\vert\Psi\rangle\over{\vert \langle \Phi\vert\Psi\rangle\vert}}\in U(1)$.
Now define their {\it relative} $SU(2)$ "(quaternionic) phase"
as ${\langle p\vert q\rangle \over {\vert\langle p\vert q\rangle \vert}}$.
Since $U(2)\sim U(1)\times SU(2)$ this convention defines a relative $U(2)$
"phase" for our entangled states. When the entangled states in question have the same reduced density matrices this $U(2)$ transformation corresponds to the possibility of the observer in the second subsystem to rotate the shared state $\vert \Psi\rangle$ to ${\vert\Phi\rangle}$, via his freedom to employ local unitary transformations.
In the general case using this definition we can compare the local unitary transformations
(corresponding to the second subsystem) of two entangled states with different 
reduced density matrices.

It is clear that these results imply many interesting applications.
Apart from  studying the generalization of our results for nonsingular $n\times n$ density matrices via the use of the anholonomy defined by Uhlmann's connection on  the trivial bundle $GL(n, {\bf C})/U(n)$, there is also the interesting possibility of studying quantum gates defined by anholonomy transformations
over the stratification manifold of entangled qudit systems.
Though some of these issues have already been partly discussed [20,22]
we hope to report some new results in a subsequent publication.

\bigskip
\centerline{\bf Acknowledgement}
\bigskip
Financial support from the Orsz\'agos Tudom\'anyos Kutat\'asi Alap (OTKA),
grant nos T032453 and T038191 is gratefully acknowledged.

\eject

\bigskip
\bigskip
\centerline{\bf REFERENCES}
\bigskip
\noindent
[1] A. Uhlmann, Rep. Math. Phys. {\bf 24}, 229 (1986)

\noindent
[2] A. Shapere and F. Wilczek (eds), {\it Geometric Phases in Physiscs}, World Scientific, Singapore 1989

\noindent
[3] J. Tidstr\"om and E. Sj\"oqvist, Phys. Rev. {\bf A67}, 032110 (2003)  

\noindent
[4] M. Ericsson, A. K. Pati, E. Sj\"oqvist, J. Br\"annlund and D. K. L. Oi,
Phys. Rev. Lett. {\bf 91}, 090405 (2003)

\noindent
[5] E. Sj\"oqvist, A. K. Pati, A. Ekert, J. S. Anandan, M. Ericsson, D. K. L. Oi, and V. Vedral, Phys. Rev. Lett. {\bf 85}, 2845 (2000)

\noindent
[6]
J. Du, P. Zou, M. Shi, L. C. Kwek, J.-W. Pan, C. H. Oh, A. Ekert, D. K. L. Oi, and
M. Ericsson, Phys. Rev. Lett {\bf 91}, 100403 (2003) 

\noindent
[7] A. Carollo, I. Fuentes-Guridi, M. Franca Santos and V. Vedral,  Phys. Rev. Lett. {\bf 90}, 160402 (2003)

\noindent
[8] A. Uhlmann, in H. -D. Doebner, V. K. Dobrev, and P. Natterman (eds), {\cal Nonlinear, Dissipative, Irreversible Quantum Systems}, World Scientific, Singapore, 296, 1995.

\noindent
[9] M. H\"ubner, Phys. Lett. {\bf A179} 226 (1993)

\noindent
[10] R. Slater, Lett. Math. Phys. {\bf 60} 123 (2002)

\noindent
[11] 
A. Ungar, Foundations of Physics {\bf 32} 1671 (2002), 

\noindent
[12] S. Hill and W. K. Wooters, Phys. Rev. Lett {\bf 80} 2245 (1997)

\noindent
[13]
J. Chen, L. Fu, A. A. Ungar and X. Zhao, Phys. Rev. {\bf A65} 024303 (2002)

\noindent
[14] P. Arrighi and C. Patricot, J. Phys. {\bf A36} L287 (2003) 

\noindent
[15] N. Mukunda, P. K. Aravind and R. Simon, J. Phys. {\bf A36} 2347 (2003)

\noindent
[16] J. G. Ratcliffe, {\it Foundations of Hyperbolic Manifolds} Springer-Verlag
 1994.
 
\noindent
[17] N. Mukunda and R. Simon, Ann. Phys. {\bf 228} 205 (1993)

\noindent
[18] H. Hopf, Math. Ann. {\bf 104} 637 (1931)

\noindent
[19] R. Mosseri and R. Dandoloff, J. Phys. {\bf A34} 10243 (2002) 

\noindent
[20] P. L\'evay, quant-ph/0306115, to appear in J. Phys. {\bf A}

\noindent
[21] B. A. Bernevig and H-D. Chen, J. Phys. {\bf A30} 8325 (2003)

\noindent
[22] J. Dittmann and G. Rudolph, J. Geometry and Physics {\bf 10} 93 (1992)

\noindent
[23] M. G. Benedict and L. Gy. Feh\'er, Phys. Rev. {\bf D39} 3194 (1989)

\noindent
[24] J. Samuel and R. Bhandari, Phys. Rev. Lett. {\bf 60} 2339 (1988),
     J. Anandan and Y. Aharonov, Phys. Rev. {\bf D38} 1863 (1988)

\noindent
[25] S. Pancharatnam, Proc. Indian Acad. Sci. {\bf A44}, 247 (1956)
\end